\documentclass[prl,aps,superscriptaddress,nofootinbib]{revtex4}
\usepackage{graphicx} 
\usepackage{epsfig}
\usepackage{dcolumn}  
\usepackage{bm}       

\def\ii{{\rm i}}  \def\Eb{{\bf E}} \def\Rb{{\bf R}} \def\rb{{\bf r}} \def\Fb{{\bf F}}
 \def\kb{{\bf k}} \def\kh{\hat{\bf k}} \def\zh{\hat{\bf z}}  
\def\kkpar0{{Q_{\rm SP}}} \def\kkparb{{\bf{Q}}}
\def\es{{\hat{\bf e}_s}} \def\ep{{\hat{\bf e}_p}} \def\lsp{{\lambda_{\rm SP}}}

\begin{document}
\title{The plasmon Talbot effect}
\author{Mark~R.~Dennis,$^1$ Nikolay~I.~Zheludev,$^2$ and F.~Javier~Garc\'{\i}a~de~Abajo$^{3*}$}
\address{$^1$School of Mathematics, University of Southampton, Southampton SO17 1BJ, UK
\\ $^2$Optoelectronics Research Center, University of Southampton, Southampton SO17 1BJ, UK
\\ $^3$Instituto de \'Optica, CSIC, Serrano 121, 28006 Madrid, Spain
\\ $^*$Corresponding author: jga@io.cfmac.csic.es}

\begin{abstract}
The plasmon analog of the self-imaging Talbot effect is described
and theoretically analyzed. Rich plasmon carpets containing hot
spots are shown to be produced by a row of periodically-spaced
surface features. A row of holes drilled in a metal film and
illuminated from the back side is discussed as a realizable
implementation of this concept. Self-images of the row are produced,
separated from the original one by distances up to several hundreds
of wavelengths in the examples under consideration. The size of the
image focal spots is close to half a wavelength and the spot
positions can be controlled by changing the incidence direction of
external illumination, suggesting the possibility of using this
effect (and its extension to non-periodic surface features) for
far-field patterning and for long-distance plasmon-based
interconnects in plasmonic circuits, energy transfer, and related
phenomena.
\end{abstract}
\maketitle

\section{Introduction}
\label{Intro}

An important aspect in the development of new branches of optics is
the study of analogues of classical optical phenomena. In the field
of surface waves on metals (surface plasmons) this approach has
successfully met with engineered nanoscale features providing
analogues of lenses and mirrors for future plasmon-based devices
\cite{BDE03,O06,ZSC06}. Such manipulation of surface plasmons is of
much interest, both from a fundamental viewpoint \cite{BDE03} and
from a view to applications \cite{O06,ZSC06}. In practice, it is
more difficult to manipulate plasmon fields than their free-space
counterparts, as they are highly sensitive to metal surface
imperfections on the scale of the skin depth ($\sim 15$ nm);
nevertheless, they have certain advantages, like their ability to
concentrate the electromagnetic field near the surface, thus
providing a route towards compact light waveguides \cite{CSS07}, or
their capacity to unveil Raman emission from single molecules
through enhancement of the local field intensity by several orders
of magnitude with respect to the incident light \cite{TJO05}. Here,
we describe and theoretically analyze the plasmon analogue to
another well-known phenomenon of classical optics, namely the
self-imaging effect discovered by Talbot in 1836 \cite{T1836} while
studying transmission gratings and arrays of holes perforated in
metal films, and later rediscovered and explained by Lord Rayleigh
\cite{R1881,P1989}. The effect is best observed through the
formation of repeated monochromatic images of a grating at various
characteristic distances of the image plane with respect to the
grating surface.

More precisely, a transversally periodic field, paraxially
propagating, {\it revives} (self-images) to its initial
configuration after the {\it Talbot distance} $\tau=2a^2/\lambda$,
where $a$ is the transverse period and $\lambda$ is the wavelength.
In a simple analytical description, we represent the grating by a
periodic function given in Fourier series form,
\begin{eqnarray}
f(x,0)=\sum_m\,\,f_m\,\,\,\exp(\ii 2\pi mx/a), \nonumber
\end{eqnarray}
where $x$ is the direction of periodicity. The monochromatic wave
function emanating from the grating towards the $y$ direction
reduces then to
\begin{eqnarray}
f(x,y)=\sum_m\,\,f_m\,\,\,\exp(\ii 2\pi mx/a)\,\,\,\exp(\ii
2\pi\zeta_my/\lambda), \label{eqa}
\end{eqnarray}
where $\zeta_m=\sqrt{1-(m\lambda/a)^2}$. The coefficients of $x$ and
$y$ in these exponential functions define a vector of magnitude
$2\pi/\lambda$, the light momentum. In the paraxial approximation
($\lambda\ll a$), the binomial expansion
\begin{eqnarray}
\zeta_m/\lambda=\frac{1}{\lambda}-\frac{m^2}{\tau}-\left(\frac{\lambda}{a}\right)^2\frac{m^4}{4\tau}-\left(\frac{\lambda}{a}\right)^4\frac{m^6}{8\tau}-\dots
\label{eq4}
\end{eqnarray}
can be truncated at the term proportional to $m^2$, equivalent to
Fresnel diffraction. This yields
\begin{eqnarray}
f(x,y)\approx\exp(\ii 2\pi y/\lambda)\,\,\sum_m\,\,f_m\,\,\,\exp(\ii
2\pi mx/a)\,\,\,\exp(-\ii 2\pi m^2y/\tau), \label{eq0}
\end{eqnarray}
from where we immediately deduce
\begin{eqnarray}
f(x,\tau)&\approx&\exp(\ii 2\pi y/\lambda)\,\,f(x,0), \label{eq00}\\
f(x,\tau/2)&\approx&\exp(\ii 2\pi y/\lambda)\,\,f(x-a/2,0).
\label{eq000}
\end{eqnarray}
The length $\tau=2a^2/\lambda$ is indeed the Talbot distance at
which the initial field self-images (except for an overall phase
that is washed away when observing intensities), while another image
is formed at $\tau/2$, laterally shifted by half a period and
leading to an alternate definition of the Talbot distance
\cite{BK96}. When $y$ is a fraction of $\tau$, the field undergoes
{\it fractional revivals}, which in the ideal case are fractal at
irrational values of $y/\tau$ \cite{BK96,LS1971_2,L1988}. This
exotic behavior is a consequence of Gauss sums arising from paraxial
propagation, which relies on the smallness of the non-paraxiality
parameter $\lambda/a$. In practice, this approximation stands only
for a finite number of $m$'s in (\ref{eqa}), but it can be
sufficient to render well-defined focal spots, as we shall see below
for self-imaging of small features.

The Talbot effect has been studied in a variety of theoretical and
experimental situations \cite{BK96,LS1971_2,L1988,TJK96,OPD06}. This
phenomenon has an analogue in Schr\"odinger evolution of quantum
mechanical wavepackets, the quantum and fractional revivals of which
have been thoroughly discussed \cite{AP1989,B96}. Although revivals
are an exact consequence of quantum mechanics, they only arise in
optics under the paraxial approximation, and deviation from
paraxiality destroys the sensitive structure of the Talbot revivals
\cite{BK96}. However, non-paraxial propagation, which only involves
a finite number of propagating waves, exhibits some good but
approximate self-imaging near the paraxial Talbot distance
\cite{NT93,STV04}.

Self-imaging is not exclusive of periodic objects. The Montgomery
effect \cite{M1967,LKJ05} describes for instance perfect image
reconstruction of aperiodic objects made of incommensurate harmonic
components $\exp\left[\ii 2\pi(m/\sqrt{|m|})x/a\right]$, leading to
replacement of $|m|$ for $m^2$ in Eq.\ (\ref{eq0}), and obviously
maintaining the property (\ref{eq00}), but not (\ref{eq000}). Recent
work on a metal film perforated by quasiperiodic hole arrays has
also revealed concentration of transmitted light intensity in hot
spots at large distances from the film \cite{paper119}, suggesting
possible extensions of the plasmon Talbot effect to aperiodic
distributions of surface features.

\section{Self-focusing of plasmon carpets on metals: the plasmon Talbot effect}

\begin{figure}
\includegraphics[width=133mm,angle=0,clip]{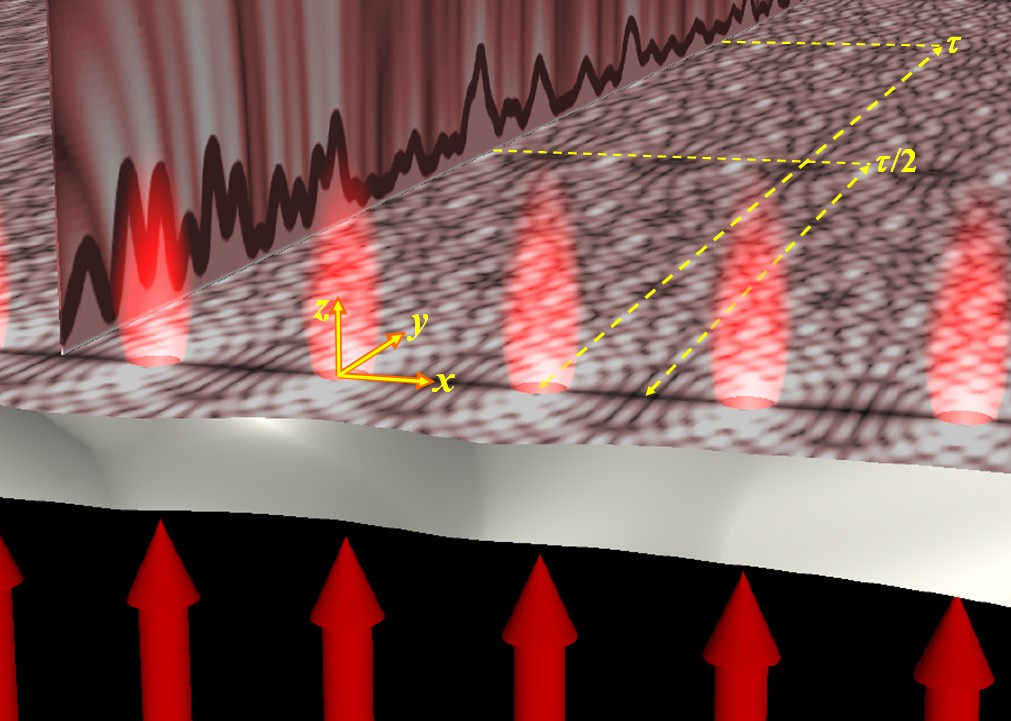}
\caption{\label{Fig1} Illustration of the plasmon Talbot effect
above a metal surface.  Light is transmitted through a
one-dimensional array of nanoholes, setting up a Talbot carpet of
interfering plasmon waves.  At approximately the Talbot distance
$\tau$ from the array, the propagating plasmons revive, giving an
array of plasmon focal spots. Plasmon revival at half that distance
is also observed, with the foci displaced by half the period along
the array direction. The dependence of the field on height $z$ above
the metal is also shown, with the intensity of the $z$ component of
the plasmons at fixed height superimposed. The carpet plotted is as
for Fig.\ \ref{Fig2}(b).}
\end{figure}

The analog of the Talbot effect using plasmons is illustrated in
Fig.\ \ref{Fig1}. A light plane wave is incident from the back of a
metal film, planar except for a periodic one-dimensional array of
nanoholes or other subwavelength structures, with period $a$. Light
is partly transmitted into plasmons on the exit side of the film,
thus deploying a complex carpet pattern. The field from each of the
nanoholes is modeled as a dipole, oscillating with a frequency
corresponding to the incident wavelength $\lambda_0$. This
oscillation sets up surface plasmons, propagating into the plasmonic
far-field with wavelength
$\lsp=\lambda_0/\Re\{\sqrt{\epsilon/(1+\epsilon)}\}$, which depends
on the particular frequency-dependent dielectric function $\epsilon$
of the metal. We shall concentrate our description on the situation
most likely to find practical application, with small attenuation
and $|\epsilon|\gg 1$, implying that $\lambda_{\rm
SP}\approx\lambda_0$. We shall also concentrate on values of the
periodicity $a$ of similar lengthscale to the plasmon wavelength
$\lambda_{\rm SP}$. In our graphical illustrations, we model a
silver surface with incident wavelength $\lambda_0=1.55\,\mu$m, for
which $\epsilon=-130.83+\ii\,3.32$ \cite{JC1972}, giving
$\lsp=1.544\,\mu$m.

Our detailed analysis starts with the field due to an oscillating
single dipole in the $y$ direction at position $\Rb_0$
infinitesimally close to the metal surface, incorporating direct
propagation and reflection.  The electric field, made dimensionless
through multiplication by $\lambda_0^3$, reads \cite{FW1984}
\begin{eqnarray}
\Eb_{\rm single}(\rb)=\int
d^2\kkparb\,\,\,\exp\left[\ii\kb\cdot(\rb-\Rb_0)\right]\,\,\,\Fb(\kkparb),
\label{eq1}
\end{eqnarray}
where
\begin{eqnarray} \Fb(\kkparb)=\frac{\ii\lambda_0^2}{Q k_z}\,\,\,\left[\ep\,k_zk_y(1-r_p)+\es\,kk_x(1+r_s)\right],
\label{eq1bis}
\end{eqnarray}
$k=2\pi/\lambda_0$ is the free-space light momentum,
$\kkparb=(k_x,k_y)$ is the projection of the wavevector $\kb$ into
the plane of the metal, $k_z=\sqrt{k^2-Q^2}$ is the component normal
to that plane, $\{\kh,\ep,\es\}$ is the natural orthonormal basis
for $\kb$, defined as $\es=\zh\times\kh/|\zh\times\kh|$ and
$\ep=\es\times\kh$, and $r_p=(\epsilon k_z-k'_z)/(\epsilon
k_z+k'_z)$ and $r_s=(k_z-k'_z)/(k_z+k'_z)$ are the appropriate
Fresnel reflection coefficients for TM ($p$) and TE ($s$)
polarization, with $k'_z=\sqrt{k^2\epsilon-Q^2}$ \cite{J99}. The
dominant component to $\Eb_{\rm single}$ is $E_z$, and this is
strongest on the metal plane for $\rb-\Rb_0$ in the direction of the
dipole (the $y$ direction). Therefore, to maximize the observable
effect, we choose to make the periodic dipole array in the $x$
direction, with the plasmons propagating in $y$.

\begin{figure}
\includegraphics[width=133mm,angle=0,clip]{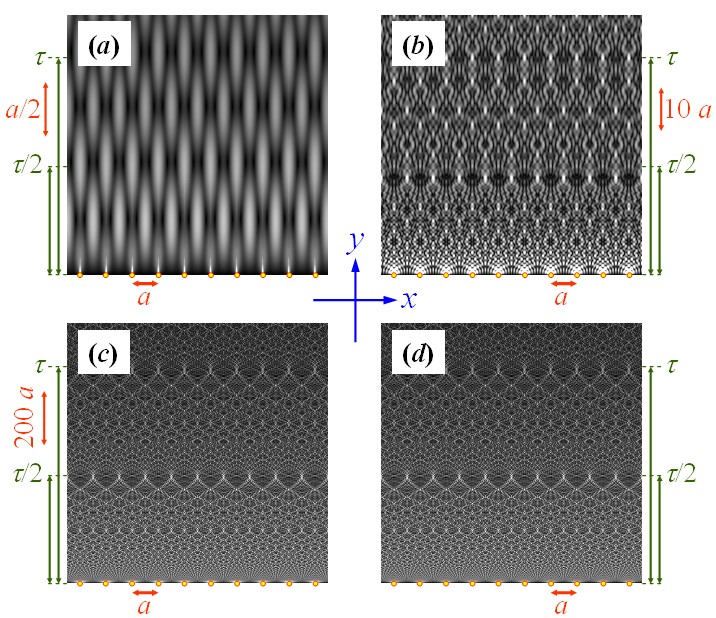}
\caption{\label{Fig2} Plasmon Talbot carpets, numerically computed
(a-c) from Eqs.\ (\ref{eq1bis})-(\ref{eq2}) and analytically
approximated (d) from Eqs.\ (\ref{eq1bis}) and (\ref{eq3}) for
different choices of the lattice spacing $a$: (a) $a=\lsp$; (b)
$a=5\,\lsp$; (c,d) $a=20\,\lsp$.  The amplitude of the $E_z$
component of the plasmon field is plotted at a height
$z=0.5\,\,\mu$m over a silver surface for a free-space wavelength
$\lambda_0=1.55\,\,\mu$m, with $\lsp=1.544\,\,\mu$m the surface
plasmon wavelength. Different scales along horizontal and vertical
directions are used in each plot: horizontal double arrows show the
period $a$, while vertical arrows signal the paraxial Talbot
distance $\tau=2a^2/\lsp$ (long arrows) and half that distance
(short arrows). The hole array is represented by circles in the
lower part of each plot. The incident light wavevector is along $z$
and its polarization along $y$ (see axes in the center of the
figure).}
\end{figure}

The ideal plasmon Talbot field comes from an infinite sum of single
dipole fields of the form of Eq.\ (\ref{eq1}), with positions at
$\Rb_n=(na,0,0)$. Using the Poisson sum formula \cite{R1970},
$\sum_n\exp(\ii k_x na)=(2\pi/a)\sum_m\delta(k_x-2\pi m/a)$, the
infinite sum can be rewritten as a Rayleigh expansion,
\begin{eqnarray}
\Eb_{\rm total}(\rb)&=&\frac{2\pi}{a}\sum_m\,\,\exp(\ii 2\pi mx/a)
\int dk_y\,\,\exp(\ii k_yy+\ii k_zz)\,\,\,\Fb(\kkparb_m) \nonumber\\
&=&\sum_m\,\,\exp(\ii 2\pi mx/a)\,\,\,\Fb_m(y,z), \label{eq2}
\end{eqnarray}
where in the first line $\kkparb_m=(2\pi m/a,k_y)$, and
$\Fb(\kkparb)$ is defined in Eq.\ (\ref{eq1bis}). In the second
line, $2\pi/a$ times the integral has been written as the $y$- and
$z$-dependent Fourier coefficient $\Fb_m(y,z)$. Numerical evaluation
of this field, for the values of the parameters above and various
choices of $a$ are shown in Fig.\ \ref{Fig2}(a-c).

For $a=\lsp$, the Talbot effect is not yet developed, although an
interesting periodic pattern appears that could be employed to
imprint hight-quality 2D arrays. When we move to larger spacing
[$a=5\lsp$ in Fig.\ \ref{Fig2}(b)], clear evidence of self-imaging
is observed, which is particularly intense at half the Talbot
distance. With even larger spacing [$a=20\lsp$ in Fig.\
\ref{Fig2}(c)] a fine Talbot carpet is deployed, showing structures
reminiscent of cusp caustics at $\tau$ and $\tau/2$ \cite{BB99}. The
focal-spot intensities decrease with distance from the hole array
due to plasmon attenuation ($\approx 1.26$ mm for silver at
$\lambda_0=1.55\,\mu$m), to which image contrast is however
insensitive at these low-absorption levels.

\begin{figure}
\includegraphics[width=130mm,angle=0,clip]{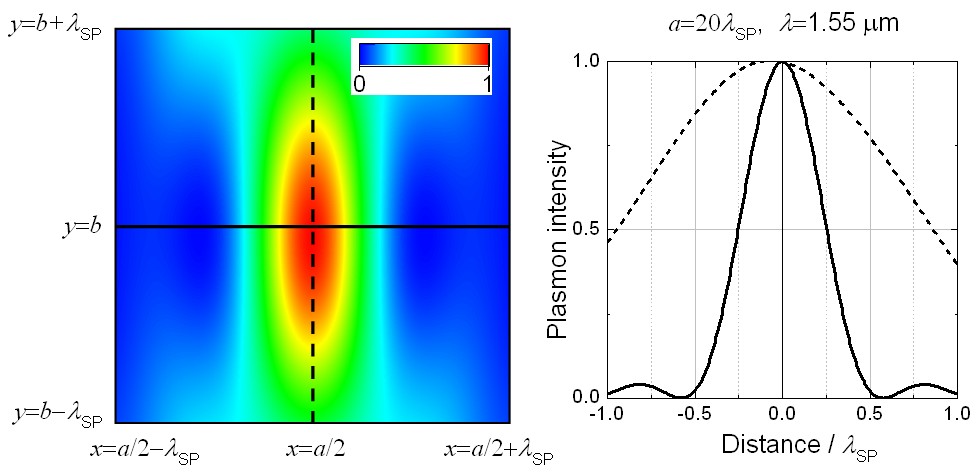}
\caption{\label{Fig3} Shape of a plasmon focal spot near half the
Talbot distance in Fig.\ \ref{Fig2}(c). The contour plot (left)
shows a square of side $2\lsp$ centered at $(x,y)=(a/2,b)$, with
$a=20\lsp$ and $b=\tau/2-5\lsp=395\lsp$. Plasmon intensities at
cross sections of the spot are given on the right along directions
parallel (solid curve) and perpendicular (broken curve) with respect
to the hole array.}
\end{figure}

The plasmon intensity in the vicinity of slightly less than half the
paraxial Talbot distance is shown in Fig.\ \ref{Fig3} for the same
conditions as in Fig.\ \ref{Fig2}(c). The plot on the left shows the
field intensity of a focal spot, with cross sectional intensities
represented on the right. The lateral width of the spot is $\approx
0.5\lsp$, whereas its extension along $y$ is considerably larger.
This type of behavior is also observed for other values of the
period and for spots at integer Talbot distances. The width along
$x$ varies from case to case, but it is always close to half a
wavelength.

\section{Analytical approach}

It is possible to approximate the field of Eq.\ (\ref{eq2})
analytically from the observation that the main contribution to the
integral over $k_y$, particularly in the plasmon far-field, comes
from the pole of the $r_p$ reflection coefficient, in the $Q$
upper-half complex plane. After all, the plasmon dispersion relation
derives from that pole (i.e., $\epsilon k_z+k'_z=0$), so that the
plasmon itself posses $p$ symmetry. This contribution may be
approximated by the Cauchy integral theorem using the $r_p$ plasmon
pole of wavenumber $\kkpar0=k\sqrt{\epsilon/(\epsilon+1)}$,
corresponding to real plasmon wavelength $\lsp=2\pi/\Re\{\kkpar0\}$
\cite{FW1984}. The remaining $z$ component of the wavevector is
$k_{z,{\rm SP}}=\sqrt{k^2-\kkpar0^2}=k/\sqrt{1+\epsilon}$. The
approximation of the integral by the pole residue is appropriate
with a cutoff on the Fourier sum in $|m|\le N$, where $N\approx
a/\lsp$. For $m$ in this range, the approximation gives
\begin{eqnarray}
\Fb_m(y,z)\approx\frac{2\lambda_0(2\pi)^3\epsilon^2}{a(\epsilon+1)^2(\epsilon-1)}
\exp\left(\frac{\ii
kz}{\sqrt{\epsilon+1}}\right)\,\exp(\ii\kkpar0\zeta_my)\,\,
\left(-m\frac{\lambda_0}{a}\frac{\sqrt{\epsilon+1}}{\epsilon},\,\,\frac{-\zeta_m}{\sqrt{\epsilon}},\,\,1\right),
\label{eq3}
\end{eqnarray}
where $\zeta_m=\sqrt{1-(2\pi m/a\kkpar0)^2}$.  This analytical
expression yields the same structure as Eq.\ (\ref{eqa}), and
therefore the general explanation of the Talbot effect offered in
Sec.\ \ref{Intro} applies here as well (assuming that the imaginary
part of $\kkpar0$ is small enough to be neglected), apart from the
extreme non-paraxiality of the regime under consideration. It should
be noted that the $m$ dependence of $\Fb_m$ is only in the vector
and in the exponent of $y$, and therefore, the Talbot carpet is
independent of $z$ in this plasmon-pole approximation, except for a
global exponential decay away from the surface.

The evaluation of Eq.\ (\ref{eq3}) corresponding to the conditions
of Fig.\ \ref{Fig2}(c) is plotted in Fig.\ \ref{Fig2}(d). Clearly,
the approximation yields excellent results, particularly in the
plasmonic far-field. However, the finite cutoff in the Fourier sum
implies that there is a finite resolution to all of the interference
features in the plasmon field, and hence a finite number of
fractional revivals (and obviously no fractal revivals), within a
Talbot length.

\begin{figure}
\includegraphics[width=130mm,angle=0,clip]{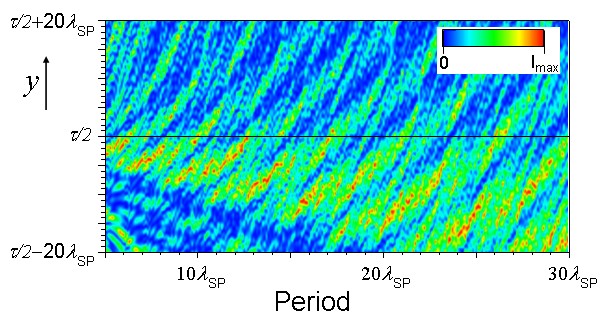}
\caption{\label{Fig4} Lattice-period dependence of the intensity
near half the Talbot distance at $x=a/2$ --in the paraxial Talbot
effect [Eq.\ (\ref{eq0})] the focal spot occurs at exactly
$y=\tau/2$. The plasmon intensity is represented along $y$ (vertical
axis) as a function of lattice period $a$ at a height
$z=0.5\,\,\mu$m over a silver surface for a free-space wavelength
$\lambda_0=1.55\,\,\mu$m. The intensity is normalized to the maximum
within the plotted range of $y$ for each period.}
\end{figure}

The choices of the periodicity $a\le 20\lsp$ in Fig.\ \ref{Fig2} are
in the non-paraxial regime. In Ref.\ \cite{BK96}, a post-paraxial
approximation to the classical Talbot effect was studied, in which
Eq.\ (\ref{eq4}) was truncated at the term proportional to $m^4$.
The inclusion of this and later terms implies that the field is no
longer perfectly periodic, and that the distance in y at which the
(imperfect) self-imaging occurs is less than $\tau$ (as in Fig.\
\ref{Fig2}). However, as our simulations and analytic approximation
demonstrate, good, if not perfect, Talbot focusing of plasmons
should nevertheless be possible in practice (similar effects have
been noticed in free-space propagation \cite{NT93,STV04}). The
dependence on the period of the focal spot near
$(x,y)=(a/2,\tau/2)$, calculated from Eq.\ (\ref{eq2}), is
illustrated in Fig.\ \ref{Fig4}, which shows a complex evolution of
the spot positions, generally below $y=\tau/2$. An interesting
consequence of these results is that the position of the focal spot
can be controlled through small changes in wavelength.

For very large values of $|\epsilon|$, electric dipoles parallel to
the metal surface are quenched by their image charges. Then, the
transmission through the holes depicted in Fig.\ \ref{Fig1} relies
on parallel magnetic dipoles (provided such dipoles can be induced,
for instance under the condition that the metal skin depth is small
compared to the hole size \cite{paper130}). Magnetic dipoles couple
best to plasmons propagating in the $y$ direction when they are
oriented along $x$. The above analysis remains valid in that case,
and in particular Eq.\ (\ref{eq3}) is only corrected by a factor
$\sqrt{\epsilon+1}$ multiplying the right-hand side. Normal electric
dipoles ($\parallel \zh$) are also relevant under these conditions,
induced by $p$-polarized light under oblique incidence. Again, Eq.\
(\ref{eq3}) can be still applied, amended by a factor
$\sqrt{\epsilon}/\zeta_m$.

\section{Discussion}

Some degree of control over the position of the hot spots is
possible when the incident light direction has non-zero projection
along the hole array direction $x$: the self-image is displaced
along $y$ from the Talbot distance and it is also laterally shifted
along $x$, as shown both theoretically and experimentally in Ref.\
\cite{TJK96}. Under these conditions, the projection of the incident
light momentum along the hole array, $k_x^i$, enters Eq.\
(\ref{eq3}) through an uninteresting overall phase factor, but also
through the coefficient of the exponential in $y$, $\kkpar0\zeta_m$,
which becomes $\sqrt{\kkpar0^2-(2\pi m/a+k_x^i)^2}$. In the paraxial
approximation, one recovers self-imaging at the corrected Talbot
distance $\tau\times(k_{y,{\rm SP}}/\kkpar0)^3$, where $k_{y,{\rm
SP}}=\sqrt{\kkpar0^2-(k_x^i)^2}$. Simultaneously, the revival is
shifted along $x$ a distance $y\,\,k_x^i/k_{y,{\rm SP}}$ that
increases with separation from the array. Thus, the position of the
focal spots can be controlled through obliquity of the external
illumination in a setup as in Fig.\ \ref{Fig1}. One should therefore
be able to raster the plasmon focus with nanometer accuracy for
potential applications in nanolithography and biosensing.

Controllable plasmon focal spots can be particularly advantageous
when combined with recently developed adaptive ultrafast nano-optics
\cite{paper120}, in which femtosecond laser pulses are shaped to
achieve a desired objective, such as a time-controlled excursion of
focal spots in the setup of Fig.\ \ref{Fig1}.

Superoscillating fields with sub-wavelength localization \cite{BP06}
should also be observed with surface plasmon waves using
appropriately designed diffraction gratings, as has been recently
observed in free-space fields generated by a quasi-crystal array of
holes \cite{paper119}.

The analysis presented here can be straightforwardly extrapolated to
other types of 2D light waves, such as guided modes in (lossless)
dielectric films, long-range surface exciton polaritons
\cite{YSB1990}, or surface modes in patterned perfect-conductor
surfaces \cite{UT1972}, with interference between metal patterns and
Talbot carpets possibly giving rise to unexpected effects in the
finer details of the surface modes. The Talbot effect is an
attribute of waves, regardless their nature, so it must occur in
sound, in elastic waves, and in the more exotic scenario offered by
electronic surface states in clean surfaces like Au(111), involving
wavelengths in the range of a few nanometers at the Fermi level
\cite{paper059} (e.g., Talbot carpets could be produced in the
vicinity of straight-line steps periodically decorated with adhered
nanoparticles).

\section{Conclusion}

We have described theoretically the surface plasmon analogue to the
classical Talbot effect.  Our numerical calculation of the dominant
normal component agrees well with our analytic approximation in the
plasmon far-field.  With weak plasmon attenuation, strong focusing
of plasmon waves is attainable, even in the non-paraxial regime that
we have studied, and some control over the position of this focusing
is possible by oblique illumination of the incident optical field.

The plasmonic Talbot effect suggests a straightforward and
implementable way of tightly focusing plasmon waves on a metal
surface. Despite the lack of perfect self-imaging imposed by the
diffraction limit, the focusing is strong enough to allow
applications in sensing and imaging. Other potential applications
include optical interconnects based upon plasmon focal spots aimed
at plasmon waveguides. We have emphasized the simplest case in which
the effect should be strongest, namely the normal component of the
field emanating from a periodic array of holes on the metal surface.
Extensions of the present work to the general case of arbitrary
distributions of holes could become an avenue to produce on-demand
plasmon fields at far distances from the holes.

\section*{Acknowledgments}

This work was supported in part by the Spanish MEC (contract No.
NAN2004-08843-C05-05), by the EPSRC (UK), and by the EU (STREP
STRP-016881-SPANS and NoE Metamorphose). MRD is supported by the
Royal Society of London.


\begin{thebibliography}{31}
\expandafter\ifx\csname
natexlab\endcsname\relax\def\natexlab#1{#1}\fi
\expandafter\ifx\csname bibnamefont\endcsname\relax
  \def\bibnamefont#1{#1}\fi
\expandafter\ifx\csname bibfnamefont\endcsname\relax
  \def\bibfnamefont#1{#1}\fi
\expandafter\ifx\csname citenamefont\endcsname\relax
  \def\citenamefont#1{#1}\fi
\expandafter\ifx\csname url\endcsname\relax
  \def\url#1{\texttt{#1}}\fi
\expandafter\ifx\csname urlprefix\endcsname\relax\def\urlprefix{URL
}\fi \providecommand{\bibinfo}[2]{#2}
\providecommand{\eprint}[2][]{\url{#2}}

\bibitem[{\citenamefont{Barnes et~al.}(2003)\citenamefont{Barnes, Dereux, and
  Ebbesen}}]{BDE03}
\bibinfo{author}{\bibfnamefont{W.~L.} \bibnamefont{Barnes}},
  \bibinfo{author}{\bibfnamefont{A.}~\bibnamefont{Dereux}}, \bibnamefont{and}
  \bibinfo{author}{\bibfnamefont{T.~W.} \bibnamefont{Ebbesen}},
  \bibinfo{journal}{Nature} \textbf{\bibinfo{volume}{424}},
  \bibinfo{pages}{824} (\bibinfo{year}{2003}).

\bibitem[{\citenamefont{Ozbay}(2006)}]{O06}
\bibinfo{author}{\bibfnamefont{E.}~\bibnamefont{Ozbay}},
  \bibinfo{journal}{Science} \textbf{\bibinfo{volume}{311}},
  \bibinfo{pages}{189} (\bibinfo{year}{2006}).

\bibitem[{\citenamefont{Zia et~al.}(2006)\citenamefont{Zia, Schuller, Chandran,
  and Brongersma}}]{ZSC06}
\bibinfo{author}{\bibfnamefont{R.}~\bibnamefont{Zia}},
  \bibinfo{author}{\bibfnamefont{J.~A.} \bibnamefont{Schuller}},
  \bibinfo{author}{\bibfnamefont{A.}~\bibnamefont{Chandran}}, \bibnamefont{and}
  \bibinfo{author}{\bibfnamefont{M.~L.} \bibnamefont{Brongersma}},
  \bibinfo{journal}{Materials\ Today} \textbf{\bibinfo{volume}{9}},
  \bibinfo{pages}{20} (\bibinfo{year}{2006}).

\bibitem[{\citenamefont{Conway et~al.}(2007)\citenamefont{Conway, Sahni, and
  Szkopek}}]{CSS07}
\bibinfo{author}{\bibfnamefont{J.~A.} \bibnamefont{Conway}},
  \bibinfo{author}{\bibfnamefont{S.}~\bibnamefont{Sahni}}, \bibnamefont{and}
  \bibinfo{author}{\bibfnamefont{T.}~\bibnamefont{Szkopek}},
  \bibinfo{journal}{Opt.\ Express} \textbf{\bibinfo{volume}{15}},
  \bibinfo{pages}{4474} (\bibinfo{year}{2007}).

\bibitem[{\citenamefont{Talley et~al.}(2005)\citenamefont{Talley, Jackson,
  Oubre, Grady, Hollars, Lane, Huser, Nordlander, and Halas}}]{TJO05}
\bibinfo{author}{\bibfnamefont{C.~E.} \bibnamefont{Talley}},
  \bibinfo{author}{\bibfnamefont{J.~B.} \bibnamefont{Jackson}},
  \bibinfo{author}{\bibfnamefont{C.}~\bibnamefont{Oubre}},
  \bibinfo{author}{\bibfnamefont{N.~K.} \bibnamefont{Grady}},
  \bibinfo{author}{\bibfnamefont{C.~W.} \bibnamefont{Hollars}},
  \bibinfo{author}{\bibfnamefont{S.~M.} \bibnamefont{Lane}},
  \bibinfo{author}{\bibfnamefont{T.~R.} \bibnamefont{Huser}},
  \bibinfo{author}{\bibfnamefont{P.}~\bibnamefont{Nordlander}},
  \bibnamefont{and} \bibinfo{author}{\bibfnamefont{N.~J.} \bibnamefont{Halas}},
  \bibinfo{journal}{Nano\ Lett.} \textbf{\bibinfo{volume}{5}},
  \bibinfo{pages}{1569} (\bibinfo{year}{2005}).

\bibitem[{\citenamefont{Talbot}(1836)}]{T1836}
\bibinfo{author}{\bibfnamefont{H.~F.} \bibnamefont{Talbot}},
  \bibinfo{journal}{Philos.\ Mag.} \textbf{\bibinfo{volume}{9}},
  \bibinfo{pages}{401} (\bibinfo{year}{1836}).

\bibitem[{\citenamefont{{Lord Rayleigh}}(1881)}]{R1881}
\bibinfo{author}{\bibnamefont{{Lord Rayleigh}}}, \bibinfo{journal}{Philos.\
  Mag.} \textbf{\bibinfo{volume}{11}}, \bibinfo{pages}{196–}
  (\bibinfo{year}{1881}).

\bibitem[{\citenamefont{Patorski}(1989)}]{P1989}
\bibinfo{author}{\bibfnamefont{K.}~\bibnamefont{Patorski}},
  \bibinfo{journal}{Prog.\ Opt.} \textbf{\bibinfo{volume}{27}},
  \bibinfo{pages}{1} (\bibinfo{year}{1989}).

\bibitem[{\citenamefont{Berry and Klein}(1996)}]{BK96}
\bibinfo{author}{\bibfnamefont{M.~V.} \bibnamefont{Berry}} \bibnamefont{and}
  \bibinfo{author}{\bibfnamefont{S.}~\bibnamefont{Klein}},
  \bibinfo{journal}{J.\ Mod.\ Opt.} \textbf{\bibinfo{volume}{43}},
  \bibinfo{pages}{2139} (\bibinfo{year}{1996}).

\bibitem[{\citenamefont{Lohmann and Silva}(1971)}]{LS1971_2}
\bibinfo{author}{\bibfnamefont{A.~W.} \bibnamefont{Lohmann}} \bibnamefont{and}
  \bibinfo{author}{\bibfnamefont{D.~E.} \bibnamefont{Silva}},
  \bibinfo{journal}{Opt.\ Commun.} \textbf{\bibinfo{volume}{2}},
  \bibinfo{pages}{413} (\bibinfo{year}{1971}).

\bibitem[{\citenamefont{Lohmann}(1988)}]{L1988}
\bibinfo{author}{\bibfnamefont{A.~W.} \bibnamefont{Lohmann}},
  \bibinfo{journal}{Optik} \textbf{\bibinfo{volume}{79}}, \bibinfo{pages}{41}
  (\bibinfo{year}{1988}).

\bibitem[{\citenamefont{Testorf et~al.}(1996)\citenamefont{Testorf, Jahns,
  Khilo, and Goncharenko}}]{TJK96}
\bibinfo{author}{\bibfnamefont{M.}~\bibnamefont{Testorf}},
  \bibinfo{author}{\bibfnamefont{J.}~\bibnamefont{Jahns}},
  \bibinfo{author}{\bibfnamefont{N.~A.} \bibnamefont{Khilo}}, \bibnamefont{and}
  \bibinfo{author}{\bibfnamefont{A.~M.} \bibnamefont{Goncharenko}},
  \bibinfo{journal}{Opt.\ Commun.} \textbf{\bibinfo{volume}{129}},
  \bibinfo{pages}{167} (\bibinfo{year}{1996}).

\bibitem[{\citenamefont{{O'Holleran} et~al.}(2006)\citenamefont{{O'Holleran},
  Padgett, and Dennis}}]{OPD06}
\bibinfo{author}{\bibfnamefont{K.}~\bibnamefont{{O'Holleran}}},
  \bibinfo{author}{\bibfnamefont{M.~J.} \bibnamefont{Padgett}},
  \bibnamefont{and} \bibinfo{author}{\bibfnamefont{M.~R.}
  \bibnamefont{Dennis}}, \bibinfo{journal}{Opt.\ Express}
  \textbf{\bibinfo{volume}{14}}, \bibinfo{pages}{3039} (\bibinfo{year}{2006}).

\bibitem[{\citenamefont{Averbukh and Perelman}(1989)}]{AP1989}
\bibinfo{author}{\bibfnamefont{I.~S.} \bibnamefont{Averbukh}} \bibnamefont{and}
  \bibinfo{author}{\bibfnamefont{N.~F.} \bibnamefont{Perelman}},
  \bibinfo{journal}{Phys.\ Lett.\ A} \textbf{\bibinfo{volume}{139}},
  \bibinfo{pages}{449} (\bibinfo{year}{1989}).

\bibitem[{\citenamefont{Berry}(1996)}]{B96}
\bibinfo{author}{\bibfnamefont{M.~V.} \bibnamefont{Berry}},
  \bibinfo{journal}{J.\ Phys.\ A} \textbf{\bibinfo{volume}{29}},
  \bibinfo{pages}{6617} (\bibinfo{year}{1996}).

\bibitem[{\citenamefont{Noponen and Turunen}(1993)}]{NT93}
\bibinfo{author}{\bibfnamefont{E.}~\bibnamefont{Noponen}} \bibnamefont{and}
  \bibinfo{author}{\bibfnamefont{J.}~\bibnamefont{Turunen}},
  \bibinfo{journal}{Opt.\ Commun.} \textbf{\bibinfo{volume}{98}},
  \bibinfo{pages}{132} (\bibinfo{year}{1993}).

\bibitem[{\citenamefont{Saastamoinen et~al.}(2004)\citenamefont{Saastamoinen,
  Tervo, Vahimaa, and Turunen}}]{STV04}
\bibinfo{author}{\bibfnamefont{T.}~\bibnamefont{Saastamoinen}},
  \bibinfo{author}{\bibfnamefont{J.}~\bibnamefont{Tervo}},
  \bibinfo{author}{\bibfnamefont{P.}~\bibnamefont{Vahimaa}}, \bibnamefont{and}
  \bibinfo{author}{\bibfnamefont{J.}~\bibnamefont{Turunen}},
  \bibinfo{journal}{J.\ Opt.\ Soc.\ Am.\ A} \textbf{\bibinfo{volume}{21}},
  \bibinfo{pages}{1424} (\bibinfo{year}{2004}).

\bibitem[{\citenamefont{Montgomery}(1967)}]{M1967}
\bibinfo{author}{\bibfnamefont{W.~D.} \bibnamefont{Montgomery}},
  \bibinfo{journal}{J.\ Opt.\ Soc.\ Am.} \textbf{\bibinfo{volume}{57}},
  \bibinfo{pages}{772} (\bibinfo{year}{1967}).

\bibitem[{\citenamefont{Lohmann et~al.}(2005)\citenamefont{Lohmann, Knuppertz,
  and Jahns}}]{LKJ05}
\bibinfo{author}{\bibfnamefont{A.~W.} \bibnamefont{Lohmann}},
  \bibinfo{author}{\bibfnamefont{H.}~\bibnamefont{Knuppertz}},
  \bibnamefont{and} \bibinfo{author}{\bibfnamefont{J.}~\bibnamefont{Jahns}},
  \bibinfo{journal}{J.\ Opt.\ Soc.\ Am.\ A} \textbf{\bibinfo{volume}{22}},
  \bibinfo{pages}{1500} (\bibinfo{year}{2005}).

\bibitem[{\citenamefont{Huang et~al.}(2007)\citenamefont{Huang, Zheludev, Chen,
  and {Garc\'{\i}a de Abajo}}}]{paper119}
\bibinfo{author}{\bibfnamefont{F.~M.} \bibnamefont{Huang}},
  \bibinfo{author}{\bibfnamefont{N.}~\bibnamefont{Zheludev}},
  \bibinfo{author}{\bibfnamefont{Y.}~\bibnamefont{Chen}}, \bibnamefont{and}
  \bibinfo{author}{\bibfnamefont{F.~J.} \bibnamefont{{Garc\'{\i}a de Abajo}}},
  \bibinfo{journal}{Appl.\ Phys.\ Lett.} \textbf{\bibinfo{volume}{90}},
  \bibinfo{pages}{091119} (\bibinfo{year}{2007}).

\bibitem[{\citenamefont{Johnson and Christy}(1972)}]{JC1972}
\bibinfo{author}{\bibfnamefont{P.~B.} \bibnamefont{Johnson}} \bibnamefont{and}
  \bibinfo{author}{\bibfnamefont{R.~W.} \bibnamefont{Christy}},
  \bibinfo{journal}{Phys.\ Rev.\ B} \textbf{\bibinfo{volume}{6}},
  \bibinfo{pages}{4370} (\bibinfo{year}{1972}).

\bibitem[{\citenamefont{Ford and Weber}(1984)}]{FW1984}
\bibinfo{author}{\bibfnamefont{G.~W.} \bibnamefont{Ford}} \bibnamefont{and}
  \bibinfo{author}{\bibfnamefont{W.~H.} \bibnamefont{Weber}},
  \bibinfo{journal}{Phys.\ Rep.} \textbf{\bibinfo{volume}{113}},
  \bibinfo{pages}{195} (\bibinfo{year}{1984}).

\bibitem[{\citenamefont{Jackson}(1999)}]{J99}
\bibinfo{author}{\bibfnamefont{J.~D.} \bibnamefont{Jackson}},
  \emph{\bibinfo{title}{Classical Electrodynamics}}
  (\bibinfo{publisher}{Wiley}, \bibinfo{address}{New York},
  \bibinfo{year}{1999}).

\bibitem[{\citenamefont{Rudin}(1941)}]{R1970}
\bibinfo{author}{\bibfnamefont{W.}~\bibnamefont{Rudin}},
  \emph{\bibinfo{title}{Real and Complex Analysis}}
  (\bibinfo{publisher}{McGraw-Hill}, \bibinfo{address}{London},
  \bibinfo{year}{1941}).

\bibitem[{\citenamefont{Berry and Bodenschatz}(1999)}]{BB99}
\bibinfo{author}{\bibfnamefont{M.~V.} \bibnamefont{Berry}} \bibnamefont{and}
  \bibinfo{author}{\bibfnamefont{E.}~\bibnamefont{Bodenschatz}},
  \bibinfo{journal}{J.\ Mod.\ Opt.} \textbf{\bibinfo{volume}{46}},
  \bibinfo{pages}{349} (\bibinfo{year}{1999}).

\bibitem[{\citenamefont{{Garc\'{\i}a de Abajo}}(in press)}]{paper130}
\bibinfo{author}{\bibfnamefont{F.~J.} \bibnamefont{{Garc\'{\i}a de Abajo}}},
  \bibinfo{journal}{Rev.\ Mod.\ Phys.}  (\bibinfo{year}{in press}).

\bibitem[{\citenamefont{Aeschlimann et~al.}(2007)\citenamefont{Aeschlimann,
  Bauer, Bayer, Brixner, {Garc\'{\i}a de Abajo}, Pfeiffer, Rohmer, Spindler,
  and Steeb}}]{paper120}
\bibinfo{author}{\bibfnamefont{M.}~\bibnamefont{Aeschlimann}},
  \bibinfo{author}{\bibfnamefont{M.}~\bibnamefont{Bauer}},
  \bibinfo{author}{\bibfnamefont{D.}~\bibnamefont{Bayer}},
  \bibinfo{author}{\bibfnamefont{T.}~\bibnamefont{Brixner}},
  \bibinfo{author}{\bibfnamefont{F.~J.} \bibnamefont{{Garc\'{\i}a de Abajo}}},
  \bibinfo{author}{\bibfnamefont{W.}~\bibnamefont{Pfeiffer}},
  \bibinfo{author}{\bibfnamefont{M.}~\bibnamefont{Rohmer}},
  \bibinfo{author}{\bibfnamefont{C.}~\bibnamefont{Spindler}}, \bibnamefont{and}
  \bibinfo{author}{\bibfnamefont{F.}~\bibnamefont{Steeb}},
  \bibinfo{journal}{Nature} \textbf{\bibinfo{volume}{446}},
  \bibinfo{pages}{301} (\bibinfo{year}{2007}).

\bibitem[{\citenamefont{Berry and Popescu}(2006)}]{BP06}
\bibinfo{author}{\bibfnamefont{M.~V.} \bibnamefont{Berry}} \bibnamefont{and}
  \bibinfo{author}{\bibfnamefont{S.}~\bibnamefont{Popescu}},
  \bibinfo{journal}{J.\ Phys.\ A} \textbf{\bibinfo{volume}{39}},
  \bibinfo{pages}{6965} (\bibinfo{year}{2006}).

\bibitem[{\citenamefont{Yang et~al.}(1990)\citenamefont{Yang, Sambles, and
  Bradberry}}]{YSB1990}
\bibinfo{author}{\bibfnamefont{F.}~\bibnamefont{Yang}},
  \bibinfo{author}{\bibfnamefont{J.~R.} \bibnamefont{Sambles}},
  \bibnamefont{and} \bibinfo{author}{\bibfnamefont{G.~W.}
  \bibnamefont{Bradberry}}, \bibinfo{journal}{Phys.\ Rev.\ Lett.}
  \textbf{\bibinfo{volume}{64}}, \bibinfo{pages}{559} (\bibinfo{year}{1990}).

\bibitem[{\citenamefont{Ulrich and Tacke}(1972)}]{UT1972}
\bibinfo{author}{\bibfnamefont{R.}~\bibnamefont{Ulrich}} \bibnamefont{and}
  \bibinfo{author}{\bibfnamefont{M.}~\bibnamefont{Tacke}},
  \bibinfo{journal}{Appl.\ Phys.\ Lett.} \textbf{\bibinfo{volume}{22}},
  \bibinfo{pages}{251} (\bibinfo{year}{1972}).

\bibitem[{\citenamefont{Mugarza et~al.}(2001)\citenamefont{Mugarza, Mascaraque,
  {V. P\'{e}rez-Dieste}, Repain, Rousset, {Garc\'{\i}a de Abajo}, and
  Ortega}}]{paper059}
\bibinfo{author}{\bibfnamefont{A.}~\bibnamefont{Mugarza}},
  \bibinfo{author}{\bibfnamefont{A.}~\bibnamefont{Mascaraque}},
  \bibinfo{author}{\bibnamefont{{V. P\'{e}rez-Dieste}}},
  \bibinfo{author}{\bibfnamefont{V.}~\bibnamefont{Repain}},
  \bibinfo{author}{\bibfnamefont{S.}~\bibnamefont{Rousset}},
  \bibinfo{author}{\bibfnamefont{F.~J.} \bibnamefont{{Garc\'{\i}a de Abajo}}},
  \bibnamefont{and} \bibinfo{author}{\bibfnamefont{J.~E.}
  \bibnamefont{Ortega}}, \bibinfo{journal}{Phys.\ Rev.\ Lett.}
  \textbf{\bibinfo{volume}{87}}, \bibinfo{pages}{107601}
  (\bibinfo{year}{2001}).

\end{thebibliography}

\end{document}